\documentclass[aps]{revtex4}
\begin{document}
\preprint{\vbox{\hbox{UJ-3252}}}

\bigskip
\bigskip

\title{The masses of $D_{sj}^\ast$(2317) and $D_{sj}^\ast$(2463) in the MIT bag model}
\author{Mariusz Sadzikowski}

\address{M. Smoluchowski Institute of Physics, Jagellonian
University, Reymonta 4, 30-059 Krak\'ow, Poland.}

\begin{abstract}
The masses of the $D_{sj}^\ast$(2317)and $D_{sj}^\ast$(2463) were found in the MIT bag model in a
reasonable agreement with the experimental values. The spectrum of ${0^\pm,1^\pm}$ for $D$ and $B$
mesons is presented. The equalities between the mass splitting $(1^+ - 1^-)(c\bar{q})\approx (1^+ -
1^-)(c\bar{s})$ and $(0^+ - 0^-)(c\bar{q})\approx (0^+ - 0^-)(c\bar{s})$, where $q=u, d$ are
confirmed despite of the fact that $SU(3)_L\times SU(3)_R$ chiral symmetry is broken by the
non-zero strange quark mass.
\end{abstract}

\pacs{PACS numbers(s):} \maketitle

\section{Introduction}

A new scalar meson with the $c\bar{s}$ flavor content and a mass around 2.32 GeV has been recently
discovered by the BaBar collaboration \cite{auber} as a narrow resonance in the $D_s\pi^0$ decay
channel and confirmed by the CLEO collaboration \cite{cleo}. This state aroused a particular
interest because of a plain conflict with the theoretical calculations which predict the mass about
$100-300$ MeV above its experimental value \cite{godfrey}. This discrepancy opened a possibility
for more refine explanations of the new state \cite{barnes}. However, it is not clear, weather the difference between the predictions and experiments rises a real issue or it is rather a problem of the models. Besides all that, there are successful explanations of the meson spectrum based on the chiral models \cite{nowak,bardeen1,feldmann,bardeen2,nowak2} under the assumption of the $SU(3)_L\times SU(3)_R$ chiral symmetry. The QCD sum rules results \cite{dai} also agree with
the experiments within the theoretical uncertainty.

This short note is, as a matter of fact, a supplement to the papers devoted to the spectroscopy of
hadrons which contain one heavy quark \cite{shuryak,izatt,sad}. In these papers the MIT bag model
\cite{chodos} was adapted to heavy-light systems ($Q\bar{q}, Qqq$ were $Q=c, b$ and $q=u, d, s$)
and masses of mesons and barions were calculated for the lowest energy states
$J^P=0^-,1^-,(\frac{1}{2})^+,(\frac{3}{2})^+$. Using the same model and the same parameters (with
justified changes in the quark masses) one can find the values of 2.29 GeV for the $0^+$ and $2.46$
for the $1^+$ $c\bar{s}$ states. 

\section{Mass calculation}

The mass of the bag can be written in the form \cite{chodos2}:
\begin{equation}
\label{m}
M = E_Q+E_q+E_{bag}+E_0+E_M+E_E
\end{equation}
where $E_{Q(q)}$ is the energy of the heavy(light) quark, $E_{bag}=\frac{4}{3}\pi B R^3$ is the
volume energy of the spherical bag of the radius $R$, $E_0=Z/R$ is the Casimir energy and the last
two terms describe the energies of the color-magnetic and color-electric interactions between
quarks. Let us first notice that for the heavy-light systems the center of mass of the bag is well
defined. It coincides with the heavy quark position and the corrections due to the c.m. motion can
be neglected. The energy of the heavy quark is approximated by the formula $E_Q=m_Q +
p^2_{light}/2m_Q$, where $m_Q$ is the mass of the heavy quark and the second term is the
non-relativistic kinetic energy. The total momentum of the bag in the c.m. frame should be zero
thus the values of the heavy and light quark momenta are equal. Let us notice that the
non-relativistic kinetic energy term in $E_Q$ was neglected in the previous calculations
\cite{izatt,sad}. This is justified if one wishes to study the lowest energy states. However for
exited states, the contribution of the kinetic energy of the heavy quark is more important and
should be included. The energy of the color-magnetic interactions between the heavy and light
quarks is equal to \cite{izatt,sad}
\begin{equation}
\label{em}
E_M=\frac{\lambda A\alpha_s}{m_QR^2}\langle\vec{\sigma}_Q\vec{\sigma}_q\rangle
\end{equation}
where $\lambda =\{1,2\}$ for barions and mesons respectively, $A$ is calculated from the light
quark wave-function and depends on its mass. The constant
\begin{equation}
\alpha_s = \frac{0.47}{\log (1+1/\Lambda R)},\,\,\,\Lambda=0.42\, \mbox{GeV}
\end{equation}
is inspired by the QCD running coupling formula \cite{sad,donoghue}. The spin-spin interaction term
$\langle\vec{\sigma}_Q\vec{\sigma}_q\rangle = \{-3,1\}$ for scalar and vector mesons respectively.
There are two sources which contribute to the color-electric energy: the interactions between
quarks and their self-interactions. For mesons with one heavy quark it can be written as
\cite{izatt}:
\begin{equation}
\label{ee} E_E=-\frac{4\alpha_s}{3R}\left(C+\frac{1}{2}\right)
\end{equation}
where $C$ can be calculated and depends on the mass of the light quark. The second term in the
parenthesis comes from the self-interaction of the heavy quark. The heavy quark self-interaction is
divergent and the infinite term is omitted in (\ref{ee})\cite{izatt}. The radius of the bag is
determined by minimizing the formula $E^\prime = E_{bag}+E_0+E_q$ with respect to $R$\footnote{One
can include the kinetic part of the heavy quark energy in $E^\prime$. The change in the radius can
be absorbed by the changes of the parameters of the model with no important impact on the spectrum.
The discussion of the interaction treatment for $R$ determination can be found in \cite{izatt}.}.

The value of the bag constant $B^{1/4}=0.1383$ and the Casimir parameter $Z=-0.73$ are taken from
the paper \cite{sad}. The masses of the quarks $m_u=m_d=0$, $m_s=0.198$ GeV, $m_c=1.824$ GeV,
$m_b=5.21$ GeV were determined to reproduce the masses of the mesons $D, D_s, B$. Using model
parameters in the $0^-,1^-$ sector one finds the light quark energy $E_q=2.04/R$ for the $u, d$
quarks and $E_s=\sqrt{m_s^{2}+(2.35)^2/R^2}$ for the strange quark. The radius of the non-strange
meson is $R=4.1$ GeV$^{-1}$ and for the strange meson $R_s=4.2$ GeV$^{-1}$. The color-magnetic
interaction parameter is $A(m_q=0)=0.52$ and $A(m_s)=0.5$ whereas for the color-electric part one
finds $C(m_q=0)=0.52$ and $C(m_s)=0.62$. These numbers together with the equations (\ref{m} -
\ref{ee}) give the spectrum written in the Table 1.
\begin{table}[t]
\caption{Masses of pseudoscalar and vector mesons.}
\begin{center}
\begin{tabular}{|c|c|c|c|c|} \hline
             & flavors    & $J^P$  & Exp. mass MeV \cite{pdg} & bag mass GeV \\ \hline
$D$          & $c\bar{q}$ & $0^-$ & 1869.3$\pm$ 0.5 & 1.868 \\
$D^\ast$     & $c\bar{q}$ & $1^-$ & 2010.0$\pm$ 0.5 & 2.008 \\
$D_s$        & $c\bar{s}$ & $0^-$ & 1968.5$\pm$ 0.6 & 1.969 \\
$D_s^{\ast}$ & $c\bar{s}$ & $1^-$ & 2112.4$\pm$ 0.7 & 2.099 \\
\hline
\end{tabular}
\begin{tabular}{|c|c|c|c|c|} \hline
             & flavors    & $J^P$  & Exp. mass MeV \cite{pdg} & bag mass GeV \\ \hline
$B$          & $b\bar{q}$ & $0^-$ & 5279.0$\pm$ 0.5 & 5.278 \\
$B^\ast$     & $b\bar{q}$ & $1^-$ & 5325.0$\pm$ 0.6 & 5.327 \\
$B_s$        & $b\bar{s}$ & $0^-$ & 5369.6$\pm$ 2.4 & 5.363 \\
$B_s^{\ast}$ & $b\bar{s}$ & $1^-$ & -               & 5.409 \\
\hline
\end{tabular}
\end{center}
\end{table}

For the scalar and pseudovector mesons, where the light quark is in the $1P_{1/2}$ state, the light
quark energy is $E_q=3.81/R$ for the $u, d$ quarks and $E_s=\sqrt{m_s^{2}+(3.91)^2/R^2}$ for the
strange quark. The radii are $R=5.09$ GeV$^{-1}$ for the non-strange and $R_s=5.08$ GeV$^{-1}$ for
the strange mesons. The color-magnetic interaction parameter is $A(m_q=0)=0.87$ and $A(m_s)=0.84$
whereas the color-electric parameter takes values $C(m_q=0)=0.88$ and $C(m_s)=0.79$. The masses of
the $0^+, 1^+$ states are given in the Table 2.
\begin{table}[t]
\caption{Masses of scalar and pseudovector mesons.}
\begin{center}
\begin{tabular}{|c|c|c|c|c|} \hline
                    & flavors    & $J^P$ & Exp. mass MeV                   & bag mass GeV \\ \hline
$D_j$               & $c\bar{q}$ & $0^+$ & 2308$\pm 30$\cite{belle}  & 2.200 \\
$D_{j}^\ast$        & $c\bar{q}$ & $1^+$ & 2427$\pm 36$\cite{belle}  & 2.383 \\
$D_{sj}^\ast$       & $c\bar{s}$ & $0^+$ & 2317.6$\pm$ 1.3\cite{auber}     & 2.288 \\
$D_{sj}^{\ast\ast}$ & $c\bar{s}$ & $1^+$ & 2463.0$\pm$ 1.7\cite{cleo}      & 2.465 \\
\hline
\end{tabular}
\begin{tabular}{|c|c|c|c|c|} \hline
                    & flavors    & $J^P$ & Exp. mass MeV$\;\;\; $& bag mass GeV \\ \hline
$B$                 & $b\bar{q}$ & $0^+$ & -                        & 5.576 \\
$B_{j}^\ast$        & $b\bar{q}$ & $1^+$ & -                        & 5.640 \\
$B_s$               & $b\bar{s}$ & $0^+$ & -                        & 5.654 \\
$B_{sj}^{\ast\ast}$ & $b\bar{s}$ & $1^+$ & -                        & 5.716 \\
\hline
\end{tabular}
\end{center}
\end{table}

The mass splitting relations $\Delta M_{1^+1^-} = \Delta M_{0^+0^-}$ \cite{nowak,bardeen1} (where
$\Delta M_{1^+1^-} = M_{1^+} - M_{1^-}$ and so on) and $\Delta M_{1^+0^+} = \Delta M_{1^-0^-}$
\cite{bardeen2,nowak2} should hold in the chiral limit. Indeed the experiment finds \cite{cleo} $\Delta
M_{1^+1^-} - \Delta M_{0^+0^-} = 1.2\pm 2.1$ MeV and $\Delta M_{1^+0^+} - \Delta M_{1^-0^-} =
1.6\pm 2.1$ MeV for the $c\bar{s}$ mesons. The MIT bag model fails to reproduce the mass splitting
equalities at this level of accuracy. This can be expected because the MIT bag model does not
handle the chiral symmetry in the appropriate way. In spite of that the model predicts that the
following relations hold (at the level of $10-20$ MeV) independently of the flavor of the light quark:
\begin{eqnarray}
\label{splitting}
\Delta M_{1^+1^-}(c\bar{q})&\approx & \Delta M_{1^+1^-}(c\bar{s}),\;\;\;
\Delta M_{0^+0^-}(c\bar{q}) \approx \Delta M_{0^+0^-}(c\bar{s})\\\nonumber
\Delta M_{1^+0^+}(c\bar{q})&\approx & \Delta M_{1^+0^+}(c\bar{s}),\;\;\;
\Delta M_{1^-0^-}(c\bar{q}) \approx \Delta M_{1^-0^-}(c\bar{s})
\end{eqnarray}
where $q=u, d$. These relations follow from the $SU(3)_L\times SU(3)_R$. However this symmetry is
broken by the non-zero, strange quark mass and it is not clear why it should work so well. Let us
stress that from the relations (\ref{splitting}) only the last one is confirmed by the experiments
at the accuracy of $1-2$ MeV.
Other relations are the assumptions based on the $SU(3)_L\times SU(3)_R$ chiral symmetry or follow
from the model calculations.

Finally, to be clear, let us notice that the considered MIT bag model joins several features of
works \cite{izatt,sad}. The parameters $B,Z$ as well as the description of the coupling $\alpha_s$
comes from the paper \cite{sad}. The treatment of the infinity in the electric heavy quark
interaction follows paper \cite{izatt}. The heavy quark kinetic term was also added. The quark
masses are adjusted anew to accommodate the changes in the model.
\section{Conclusions}
The result of the previous paragraph, at the face value, can be regarded as a success of the MIT
bag model. The description of the meson spectrum(similar for barions \cite{shuryak,izatt,sad}) at
the level of 4-5 per cent is already at the edge of the reliability of the model.

However the more careful scrutiny leads to additional questions. Let us first notice that the quark
masses are not treated on an equal footing. The $m_u, m_d, m_s$ are current masses whereas $m_c,
m_b$ are constituent masses. This follows from the treatment of the heavy quark  self-interaction
in the color-electric energy. The infinite term is absorbed to the mass of the heavy quark.
Secondly, the coupling constant at the energy scale $1/R=0.25-0.1$ GeV takes values $\alpha_s
=1.0-1.2$. This is certainly not a small parameter. Indeed one can try to fit a new set of the bag
model parameters with current quark masses and a small coupling constant. First, the infinite term
could be cut-off by the heavy quark Compton wave-length $1/m_Q$. Secondly, the scale of the
interaction between the light and heavy quarks may be rather set by the $m_Q$ or geometrical mean
$\sqrt{m_QE_q}$. If done, then the coupling constant can be reduced to the value of $\alpha_s\sim
0.3-0.4$ and the heavy quarks get their current masses. But in this situation, the mass of
$D^{\ast}_{sj}(2317)$ immediately falls in the range $2.44-2.55$ GeV. The simultaneous requirement
of the current heavy quark masses and a small coupling constant can not be reconciled with a good
fit to the whole meson spectrum\footnote{For example $B^{1/4}=0.1383$ GeV, $Z=-1.85$, $m_u=m_d=0$,
$m_s=0.143$ GeV, $m_c=1.482$ GeV, $m_b=4.6$ GeV gives a relatively good fit to the lowest energy
hadron spectrum. The energy scale for the heavy-light quarks interactions is set to
$\sqrt{m_QE_q}$. The infinite term is regularized by the Compton wave-lenght of the heavy quark.
The author tried different sets of parameters with reasonable assumptions about the energy scale of
the interactions and regularization methods with similar success.}.

Despite of these drawbacks the MIT bag model is capable to predict the $0^\pm,1^\pm$, $D$ mesons
spectrum with the accuracy of few per cent. One can also find  that the mass splitting relations
(\ref{splitting}) are independent of the flavor of the light quark. These relations follow directly
from the chiral symmetry $SU(3)_L\times SU(3)_R$. However this symmetry is broken by the strange
quark mass and it is not obvious why the relations (\ref{splitting}) should hold.

In the further work one can try to include the light quark in the $1P_{3/2}$ state. However the
calculation of the $1P_{3/2}$ state is more involved and require reconsideration of the model
assumptions (e.g. sphericity of the bag) thus demanding much more extensive elaboration then this
short note assumed.

\bigskip

{\bf Acknowledgements}. I would like to thank Leszek Motyka and
Kacper Zalewski for useful discussions. This work  was supported
by the Polish State Committee for Scientific Research, (KBN) grant
no.  2 P03B 09322.

\end{document}